\documentclass[aps,pra,twocolumn,floatfix]{revtex4-1}

\usepackage{bm,dcolumn,amsmath,graphicx}
\usepackage{epsfig}
\usepackage{nicefrac}
\usepackage{longtable}

\def\adndt{At. Data Nucl. Data Tables}
\def\cjp{Can. J. Phys.}

\def\epl{Europhys. Lett.}

\def\jpb{J. Phys. B}

\def\jpcs{J. Phys.: Conf. Ser.}

\def\mnras{Mon. Not. R. Astron. Soc.}
\def\nat{Nature}

\def\pr{Phys. Rev.}
\def\pra{Phys. Rev. A}

\def\prl{Phys. Rev. Lett.}

\def\pscr{Phys. Scr.}
\def\ptrsa{Phil. Trans. R. Soc. A}

\def\sci{Science}

\newcommand{\paper}{paper}

\newcommand{\Eref}[1]{Eq.~(\ref{#1})}
\newcommand{\Tref}[1]{Table~\ref{#1}}
\newcommand{\Fig}[1]{Fig.~\ref{#1}}
\newcommand{\Sec}[1]{Sec.~\ref{#1}}

\newcommand{\cm}{\ensuremath{\textrm{cm}^{-1}}}

\newcolumntype{b}{D{(}{\ (}{-1}}  

\newcommand{\Zi}{\ensuremath{Z_\textrm{ion}}}
\newcommand{\Za}{\ensuremath{Z_a}}

\begin{document}

\title{Highly charged ions with E1, M1, and E2 transitions within laser range}

\author{J. C. Berengut}
\author{V. A. Dzuba}
\author{V. V. Flambaum}
\author{A. Ong}
\affiliation{School of Physics, University of New South Wales, Sydney, NSW 2052, Australia}

\date{4 June 2012}


\begin{abstract}
Level crossings in the ground state of ions occur when the nuclear charge $Z$ and ion charge \Zi\ are varied along an isoelectronic sequence until the two outermost shells are nearly degenerate. We examine all available level crossings in the periodic table for both near neutral ions and highly charged ions (HCIs). Normal E1 transitions in HCIs are in X-ray range, however level crossings allow for optical electromagnetic transitions that could form the reference transition for high accuracy atomic clocks. Optical E1 (due to configuration mixing), M1 and E2 transitions are available in HCIs near level crossings. We present scaling laws for energies and amplitudes that allow us to make simple estimates of systematic effects of relevance to atomic clocks. HCI clocks could have some advantages over existing optical clocks because certain systematic effects are reduced, for example they can have much smaller thermal shifts. Other effects such as fine-structure and hyperfine splitting are much larger in HCIs, which can allow for richer spectra. HCIs are excellent candidates for probing variations in the fine-structure constant, $\alpha$, in atomic systems as there are transitions with the highest sensitivity to $\alpha$-variation.
\end{abstract}

\maketitle

\section{Introduction}

Current technological plans hint to mainstream adoption of highly charged ions (HCIs) for many uses in the near future~(see, e.g. the review~\cite{gillaspy01jpb}). Production of any ion stage of practically any naturally occurring element is possible at ion accelerators and/or electron beam ion traps. Furthermore great progress has been made recently in trapping, cooling, and spectroscopy of HCIs (see, e.g.~\cite{draganic03prl,crespo08cjp,hobein11prl,mackel11prl} and the review~\cite{beiersdorfer09pscr}). In this paper we consider candidate transitions for an optical clock made using a HCI that has a configuration crossing in the ground state: a ``level crossing''.

Level crossings in ions occur when the energy ordering of orbitals changes with increasing ion charge. The ion charge may be increased by considering ionisation along an isonuclear sequence or by considering an isoelectronic sequence with variable nuclear charge. The latter is somewhat simpler to deal with theoretically since the electronic structure does not usually change very much between adjacent ions. In this paper we discuss isoelectronic sequences at points where the electronic structure does change --- the level crossings --- and interesting properties can emerge. Near level crossings, the frequencies of transitions involving the crossing orbitals can be much smaller than the ionisation energy. This means that they can be within the optical range and have the potential to be excited by lasers, opening the possibility of performing high-precision spectroscopy and building optical clocks using HCI reference transitions.

This work is also motivated by astronomical observations of quasar absorption spectra that hint that there is a spatial gradient in the value of the fine-structure constant, $\alpha = e^2/\hbar c$~\cite{webb11prl,king12mnras}. Data samples from the Very Large Telescope and Keck Telescope~\cite{webb99prl,murphy03mnras} independently agree on the direction and the magnitude of this gradient, which is significant at a $4.2\sigma$ level. A consequence of the astronomical result is that since the solar system is moving along this spatial gradient, there may exist a corresponding temporal shift in $\alpha$ in the Earth frame at the level $\dot\alpha/\alpha\sim10^{-19}\ \textrm{yr}^{-1}$~\cite{berengut12epl}. Finding this variation using atomic clocks could independently corroborate the astronomical result in the laboratory.

The best current terrestrial limit on time-variation of $\alpha$ was obtained by comparing the ratio of frequencies of the Al$^{+}$ clock and the Hg$^{+}$ clock over the course of a year~\cite{rosenband08sci}. The ratio is sensitive to $\alpha$-variation because the reference transitions in the two clocks have different sensitivity coefficients, $q$, defined as
\begin{equation}
\label{eq:q_def}
q = \frac{d\omega}{dx}\bigg|_{x=0}\ ,
\end{equation}
where $x = \alpha^2/\alpha^2_0 - 1$ is a normalised change in $\alpha^2$ from the current value $\alpha_0^2$, and $q$ and $\omega$ are measured in atomic units of energy. In this experiment the Al$^{+}$ clock is relatively insensitive to $\alpha$ variation (low $q$ coefficient), thus serving as an ``anchor'' line. On the other hand the Hg$^+$ clock is sensitive to $\alpha$-variation (high $q$ coefficient). Therefore, the ratio of these transition frequencies will change if $\alpha$ changes. The limit on the rate of change of $\alpha$ was measured as $\dot{\alpha}/\alpha = (-1.6\pm 2.3)\times 10^{-17}~\text{yr}^{-1}$.

To compete with astrophysical measurements of the spatial gradient, the atomic clock limits must be improved by around two orders of magnitude. Several proposals have been made for atomic clocks that, if measured at the same level of accuracy as the Al$^+$/Hg$^+$ ratio, would give much stronger limits on $\alpha$-variation. These include: proposals to construct clocks using heavier elements with similar properties (e.g. the Tl$^{+}$ clock proposed by~\cite{dehmelt89pnas}); systems with large relative sensitivities to $\alpha$-variation exploiting the accidentally degenerate levels in Dy~\cite{dzuba99prl,cingoz07prl} or fine-structure anomalies in Te, Po, and Ce~\cite{dzuba05pra}; a variety of transitions in heavy elements with large $q$ values (e.g.~\cite{dzuba03pra,angstmann04pra0,porsev09pra,flambaum09pra}); and nuclear clocks based on the 7.6eV isomeric transition in the $^{229}$Th nucleus that would have extraordinary sensitivity to variation of fundamental constants~\cite{peik03epl,flambaum06prl,berengut09prl,campbell12prl}. For a more complete review see~\cite{dzuba09cjp,berengut11jpcs}.

Transitions near level crossings in HCIs can provide higher sensitivity to $\alpha$-variation than any other optical transitions seen in atomic systems~\cite{berengut10prl,berengut11prl}. Consider the following analytical formula for the relativistic shift of an energy level in the single-particle approximation~\cite{dzuba99prl}:
\begin{equation}
\label{eq:q_approx}
q_n \approx -I_n \frac{(Z\alpha)^2}{\nu(j+1/2)} \,,
\end{equation}
where $I_n$ is the ionisation energy of the orbital (atomic units $\hbar = e = m_e = 1$), and $\nu$ is the effective principal quantum number. A transition in a HCI can have a large sensitivity because the difference in $q_n$ between the levels involved can be large. The enhancement comes from the coherent contributions of three factors: high nuclear charge $Z$, high ionisation degree $Z_\textrm{ion}$ (leading to large $I_n$), and significant differences in the configuration composition of the states involved (large changes in $j$ and $\nu$). For nearly-filled shells, an additional enhancement in the $\alpha$-sensitivity occurs due to each electron spending approximately half its time nearer to the nucleus than other electrons in the same shell. In these cases $q_n \sim I_n^{3/2}$~\cite{berengut11prl}.

In this paper we perform a systematic search for level crossings in HCIs throughout the periodic table. We identify several ranges of $Z$ and $Z_\textrm{ion}$ where level crossings can be found, and perform configuration interaction calculations for some of the most promising systems. In \Sec{sec:scalings} we discuss how systematic effects that affect optical clocks are modified in the case of HCIs, and find that HCIs confer some benefits over near-neutral ions. Current experimental techniques might be applied to build a similar clock retaining high precision, but with much higher sensitivity to $\alpha$-variation.

\section{Method}

\begin{figure}
\caption{A comparison of the ordering of electron orbitals: The first column is the order of filling as derived by applying the Madelung rule, while the second column is derived for a hydrogen-like atom (excluding $g$-wave and $h$-wave orbitals that cannot be occupied in the ground state of any real ion). The ordering of orbitals changes with increasing ion charge, $Z_{ion}$.}
\label{fig:ordering}
\includegraphics{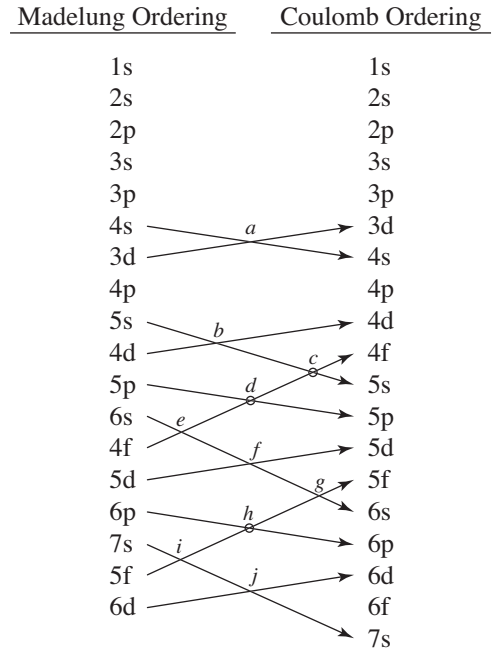}
\end{figure}

Our first task in this work is to identify HCIs with level crossings in the ground state. We start with neutral ions and then increase $Z$, working along the isoelectronic sequence from the neutral atom filling order towards the Coulomb filling order. The Madelung rule (also known as the Klechkowski rule) can be taken as a first approximation for determining the filling order of electron shells in neutral atoms. We show in the Appendix that this is a good approximation because deviations from this filling order in neutral atoms disappear with a small increase in the ion charge $Z_\textrm{ion}$. Also we know that in very highly-charged ions, the energy levels of the electrons must approach the hydrogen-like (Coulomb) limit, where all orbitals with the same principal quantum number $n$ are nearly degenerate. \Fig{fig:ordering} presents the order of electron orbitals under both ordering schemes. Since any difference in the ordering as computed from the Madelung rule and that of the hydrogen-like limit must be resolved with increasing \Zi, the `out-of-order' levels must cross at some \Zi. From the transition between these limits it is seen that the only types of crossings available in HCIs are between orbitals with angular momenta $s-d$, $s-f$, and $p-f$.

Neutral atoms sometimes have ground state electronic configurations that deviate from the Madelung rule. In isoelectronic sequences starting with such atoms other types of level crossings can occur (namely, $5d-4f$ and $6d-5f$). However, we find that the new crossings occur with the addition of just a few extra protons; no additional crossings are found in highly-charged ($\Zi \gtrsim 5$) ions. Full details are presented in the Appendix.

\begin{figure}[tb]
\caption{Dirac-Fock energies of the $4f_{5/2}$ (solid), $5p_{1/2}$ (dashed), and $5p_{3/2}$ (dotted) levels of the In ($N=49$) isoelectronic sequence.}
\label{fig:4f5p_crossing}
\includegraphics[width=0.45\textwidth]{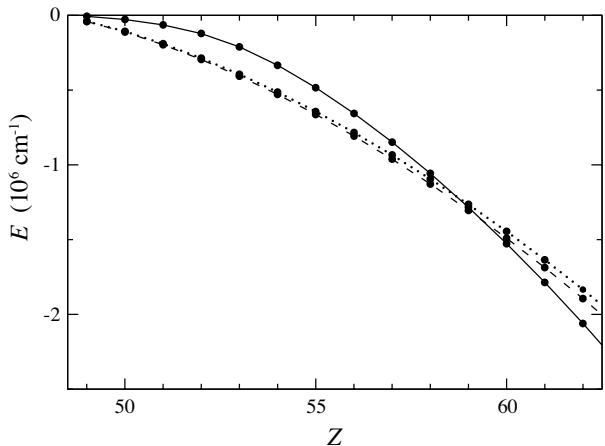}
\end{figure}

To find the ions along the isoelectronic sequence where level crossings lead to small transition frequencies we perform Dirac-Fock (relativistic Hartree-Fock) calculations. An example is presented in \Fig{fig:4f5p_crossing}, which shows the $4f$ and $5p$ valence orbitals of the indium isoelectronic sequence ($N=49$) calculated using Dirac-Fock in the $V^{N-1}$ approximation. It is seen that for low values of \Zi\ the $4f$ orbitals lie above the $5p$ orbitals, but at $Z=59$ the $4f$ levels drop below the $5p_{3/2}$ orbital, and between $Z=59$ and 60 they cross the $5p_{1/2}$ orbital energy. In general, this method produces acceptable estimates for the position of the crossings as we will see by comparison with Configuration Interaction calculations in Sections~\ref{sec:fp_crossingOne} and \ref{sec:fp_crossingTwo}.

\begin{figure}[tb]
\caption{Energies of the $4f_{5/2}$ (solid), $5p_{1/2}$ (dashed), and $5p_{3/2}$ (dotted) levels of the In ($N=49$) isoelectronic sequence. Upper panel: detail of the level crossing in the Dirac-Fock approximation of \Fig{fig:4f5p_crossing}. Lower panel: the same level crossing calculated with many-body perturbation theory corrections included. The qualitative nature of the crossing point is not significantly affected by the MBPT corrections.}
\label{fig:MBPTComp}
\includegraphics[width=0.45\textwidth]{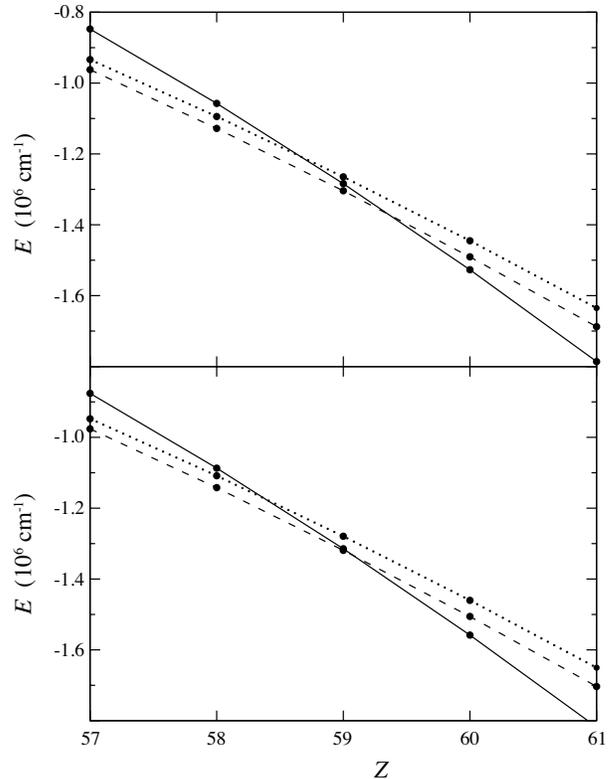}
\end{figure}

Many-body perturbation theory (MBPT) corrections can be included, but our calculations show that this does not change the position of the crossing point (see \Fig{fig:MBPTComp}, which shows a detailed view of the crossing point in \Fig{fig:4f5p_crossing} with and without including of MBPT corrections).

The indium sequence described above has one valence electron above closed shells (a cadmium core that we may consider frozen). In general, however, we can perform Dirac-Fock calculations even for several-valence-electron ions provided we scale the contribution of each subshell by its filling fraction. Again this gives reasonable accuracy for the ionisation energy (order of few percent) which is good enough to identify level crossings. As we progress along an isoelectronic sequence, we increase $Z$ until the first crossing point is reached. After this point the electronic configuration will be altered, and to find other crossing points that occur later in the sequence further calculations must be performed with the modified electron configuration which assumes that the first crossing has occurred.

In principle, it is possible to use the weighted Dirac Fock method outlined above for an arbitrary number of electrons, but for partially-filled shells and electron-hole calculations there usually will be more than one possible DF electron configuration to use. One such example is Cr II~\cite{berengut11pra1}, where the $d$-shell electrons must be accounted for in the DF approximation, but it is not clear if a $V^N$ scheme where $3d^5$ is included in the DF potential or a $V^{N-1}$ scheme where $3d^4$ is included will give better agreement with experiment (of course, in the limit of a complete basis set both approximations will give the same CI result). Furthermore, using a poor approximation for the DF potential may result in the Dirac-Fock calculation showing no available level crossings. In order to resolve this issue, we must perform at least minimal configuration interaction calculations to locate the crossing point and calculate approximate transition frequencies.

HCIs with many valence electrons have some benefits for potential clock applications because of the availability of different angular momentum states and configurations. This is useful both for finding  reference transitions with desirable properties and also for increasing the sensitivity of the transition to $\alpha$-variation, as the $q$ values for an $k$-electron transition is approximately $k$ times the $q$ value for the single electron transition. This is illustrated, e.g., by the examples presented in~\cite{berengut10prl,berengut11prl}. Furthermore, using configuration mixing it is also possible to generate E1 transitions using multiple electrons in a $s-f$ crossing. In the following sections, we will list all the available level crossings in elements from the considerations discussed above.

The calculations presented in this paper use the atomic structure code AMBiT~\cite{berengut06pra}, which includes Dirac Fock (DF) and Configuration Interaction (CI) algorithms. While core-valence calculations can be included in a CI calculation via many-body perturbation theory using the CI+MBPT method~\cite{dzuba96pra}, for our current purposes this is not required, as discussed below. All CI calculations for two and three-valence-electron ions are performed using a fairly small $B$-spline basis of the type developed in~\cite{johnson86prl,johnson88pra}, including valence orbitals only up to $7spdf$. MBPT corrections are much more important for the calculation of transition frequencies, $\omega$. More precise studies of those HCIs that are of interest to experimentalists will need to be performed using the full CI+MBPT theory.


In \Tref{tab:Licomp} we compare experimental ionization energies with those calculated in the DF approximation for several levels of neutral lithium. For this simple case, we see that the ionization energies and intervals are accurate to $\sim1\%$ or better. \Tref{tab:Wcomp} compares the ground state ionization energies for selected ions along the tungsten ($Z = 74$) isonuclear sequence with available data. In HCIs, we see that we maintain roughly the same degree of accuracy, therefore, we can surmise that the position of level crossings are fairly accurately determined from the DF calculations alone.

On the other hand, we note that the energy of transitions between the levels participating in the optical level crossing may not be as easily determined for HCIs (indeed even the ordering may be difficult to determine). This is because we are selecting HCIs where the difference between the ionization energies of these levels are strongly suppressed. For the typical scale of ionization energies in HCIs, $\sim 10^7~\textrm{cm}^{-1}$, an interval of $\sim 10^4~\textrm{cm}^{-1}$ is the result of a cancelation at the level 99.9\%. To determine the ground state conclusively an accuracy of better than 0.1\% in the ionisation energy is required. As a result, in HCIs near level crossings the electronic structure is not as well determined as in near neutral ions, and the ground state is typically not identified to a high degree of confidence. For experimental purposes however, the two (or more) possible ground states in HCIs with optical level crossings can all be considered metastable, as they have typical lifetimes ranging from seconds to the lifetime of the universe.

\begin{table}
\caption{Dirac-Fock calculation of ionization energy and energy intervals for neutral lithium, compared with available experimental data from~\cite{NIST}.}
\label{tab:Licomp}
\begin{ruledtabular}
\begin{tabular}{lcrrr}
\multicolumn{1}{c}{Level} & \multicolumn{1}{c}{$J$} & \multicolumn{2}{c}{Ionization Energy~(cm$^{-1}$)} & \multicolumn{1}{c}{\% Deviation} \\
&& \multicolumn{1}{r}{DF Calc.} & \multicolumn{1}{r}{Expt.} & \\
\hline
$2s$ & 1/2 & -43087 & -43487 & -0.919\\
$2p$ & 1/2 & -28232 & -28583 & -1.226\\
     & 3/2 & -28232 & -28583 & -1.227\\
$3s$ & 1/2 & -16197 & -16281 & -0.513\\
$3p$ & 1/2 & -12460 & -12561 & -0.808\\
     & 3/2 & -12459 & -12561 & -0.809\\
$3d$ & 3/2 & -12194 & -12204 & -0.079\\
     & 5/2 & -12194 & -12204 & -0.079\\
$4s$ & 1/2 &  -8444 &  -8475 & -0.367\\
$4p$ & 1/2 &  -6975 &  -7017 & -0.604\\
     & 3/2 &  -6974 &  -7017 & -0.604\\
$4d$ & 3/2 &  -6859 &  -6863 & -0.065\\
     & 5/2 &  -6859 &  -6863 & -0.065
\end{tabular}
\end{ruledtabular}
\end{table}

\begin{table}
\caption{Dirac-Fock calculation of energy levels for selected ions belonging to the ionization sequence of tungsten, compared with available data.}
\label{tab:Wcomp}
\begin{ruledtabular}
\begin{tabular}{lrrr}
\multicolumn{1}{c}{Ion} & \multicolumn{2}{c}{Ionization Energy~($10^3$~cm$^{-1}$)} & \multicolumn{1}{c}{\% Deviation} \\
& \multicolumn{1}{r}{DF Calc.} & \multicolumn{1}{r}{Expt.~\cite{kramida09adndt}} & \\
\hline
W$^{5+}$  &    -509 &    -522 & 2.49\\
W$^{11+}$ &   -1846 &   -1868 & 1.17\\
W$^{13+}$ &   -2440 &   -2345 & 4.05\\
W$^{27+}$ &   -7075 &   -7109 & 0.47\\
W$^{37+}$ &  -13049 &  -13080 & 0.23\\
W$^{45+}$ &  -19487 &  -19471 & 0.08\\
W$^{55+}$ &  -43101 &  -43133 & 0.07\\
W$^{73+}$ & -652346 & -651338 & 0.15
\end{tabular}
\end{ruledtabular}
\end{table}

\section{Ground State Level Crossings}

In this section we list all possible level crossings that occur because of the transition from the Madelung filling scheme to the Coulomb degenerate scheme. All crossings in \Fig{fig:ordering} are represented, however most occur at relatively low ion stage or outside the range of relatively stable nuclei ($Z\gtrsim100$). The most interesting cases are those that occur in HCIs with $\Zi\gtrsim5$: crossings $c$ ($4f-5s$), $d$ ($4f - 5p$), and $h$ ($5f - 6p$), which are studied in further detail in Sections~\ref{sec:fs_crossing}, \ref{sec:fp_crossingOne} and \ref{sec:fp_crossingTwo}, respectively. Once an ion is found with orbitals near a particular level crossing, nearby ions with the same crossing can generally be found by increasing the nuclear charge while simultaneously increasing the number of electrons by the same amount, provided that the orbital shells involved in the crossing are not completely filled.

\paragraph{$3d - 4s$:}
The earliest crossing possible in the periodic table occurs in the K isoelectronic sequence ($N=19$). The ground state configuration is [Kr]$4s$, but the ground state of Sc$^{2+}$ ($Z = 21$) is [Kr]$3d$. This crossing can be seen in the early transition metals, where it is well known that the $3d$ and $4s$ orbitals are nearly degenerate in neutral and near-neutral ions of these elements. All isoelectronic sequences beginning from neutral atoms with $19 \leq N \leq 28$ have this crossing. The $N=29$ isoelectronic sequence starts with Cu, where the ground state is $3d^{10}4s$; this sequence has no crossing since in the neutral atom the electron shells already fill in the Coulomb-limit order.

\paragraph{$4d - 5s$:}
For the Rb isoelectronic sequence, this crossing point occurs near $Z = 39$, which is Y$^{2+}$. Again this level crossing happens in near-neutral systems; it is available in isoelectronic sequences with $37 \leq N \leq 46$. For $N = 47$ the ground state already has Coulomb degenerate ordering. One ion with this crossing, the two-valence-electron ion Zr$^{2+}$, was discussed in~\cite{berengut11pra0}.

\paragraph{$4f - 5s$:}
The $5s$ and $4f$ level crossing occurs at a higher degree of ionisation than the previous two crossings. The lightest ions with this crossing occur in the $N=47$ isoelectronic sequence, which has a single electron above closed shells. The ions Nd$^{13+}$, Pm$^{14+}$ and Sm$^{15+}$ have optical transitions between these orbitals; they were studied in~\cite{berengut10prl}. The heaviest ions with this crossing occur when the $4f$ and $5s$ shells are nearly filled, i.e. in the isoelectronic sequences of Pm or Nd. These were studied in~\cite{berengut11prl} where the ions Ir$^{16+}$ and Ir$^{17+}$ (ground state configurations $4f^{13}5s^2$ and $4f^{13}5s$, respectively) were found to have optical transitions from the ground state with the extremely large $q$ values. The total number of ions with this crossing is around fifty. This level crossing is available in isoelectronic sequences with $47 \leq N \leq 61$. We discuss other examples with this crossing in \Sec{sec:fs_crossing}.

\paragraph{$4f - 5p$:}
The $5p_{1/2}$ and $5p_{3/2}$ orbitals are separated by a large fine-structure interval, which causes this level crossing to occur over a wider range of $Z$ (see \Fig{fig:4f5p_crossing}). For a single electron above a closed shell, this crossing occurs at around $Z = 59$. \Fig{fig:MBPTComp} illustrates the effect of including MBPT corrections on the position of the level crossing. This level crossing is available in isoelectronic sequences with $49 \leq N \leq 67$. The ions W$^{7+}$ and W$^{8+}$ ($N=67$ and 66, respectively), which have hole transitions between the nearly filled shells, were studied in detail in~\cite{berengut11prl}. We discuss other examples in \Sec{sec:fp_crossingOne}.

\paragraph{$4f - 6s$:}
This crossing point occurs much earlier in the ionization sequence than other $s - f$ crossings presented here since the difference in principal quantum number between the orbitals is $\Delta n = 2$. In the Cs isoelectronic sequence, this crossing occurs in Ce$^{3+}$, however the $5d$ orbital also plays a role here and the $4f-6s$ level crossing is not seen in the ground state of this sequence. This level crossing is available in isoelectronic sequences with $55 \leq N \leq 69$.

\paragraph{$5d - 6s$:}
Just as in the $4d-5s$ case, the $5d$ and $6s$ orbitals cross at low ionization stage; for the Cs isoelectronic sequence ($N=55$) it occurs in doubly-ionised lanthanum ($Z=57$). On the other hand $s^2-d^2$ transitions can have reasonably large $q$-values even in ions with relatively small ion stage, especially where the hole transitions are used. Several interesting examples including Hf$^{2+}$, Hg$^{2+}$, and Hg$^{3+}$ were studied in~\cite{berengut11pra0}. This level crossing is available in isoelectronic sequences with $55 \leq N \leq 78$.

\paragraph{$5f - 6s$:}
This crossing is similar to the $4f-5s$ crossing previously discussed, and it was hoped that ions which showed this crossing would have very high $q$-values due to the large $Z^2$ enhancement factor. However, the $6s$ orbital is much more tightly bound than the $5f$ orbitals and as a result the level crossing occurs at $Z = 105$ for the Au isoelectronic sequence and well beyond $105$ for the Tl isoelectronic sequence. While this level crossing occurs in isoelectronic sequences with $79 \leq N \leq 101$, it is unavailable in any stable nuclei.

\paragraph{$5f - 6p$:}
The $6p_{1/2}$ and $6p_{3/2}$ orbitals are very far apart in HCIs due to large fine-structure splitting (the $5f_{5/2}$ and $5f_{7/2}$ orbitals are much closer). This causes a bifurcation of this level crossing, with $5f$ crossing the $6p_{3/2}$ orbitals (in the excited state) near $Z = 93$ and $6p_{1/2}$ near $Z = 98$ for the Tl isoelectronic sequence. This level crossing is available in isoelectronic sequences with $81 \leq Z \leq 101$. It was originally exploited in~\cite{berengut12arxiv1} where it was shown that optical transitions in Cf$^{16+}$ ($N=82$ with two valence electrons) have the largest sensitivity to variation of the fine-structure constant seen in any atomic system. We discuss more examples in \Sec{sec:fp_crossingTwo}.

\paragraph{$5f - 7s$:}
As in the case of the $4f - 6s$ crossing, the difference in principal quantum number is $\Delta n = 2$. Since states with larger $n$ tend to have lower orbital energy, this causes the $7s$ orbital to be comparable in energy to $5f$, thus creating a crossing point early in the ionization sequence. Ac$^{2+}$, which is near this level crossing, was examined in~\cite{berengut11pra0}. This level crossing is available in isoelectronic sequences with $87 \leq N \leq 101$.

\paragraph{$6d - 7s$ and $6f - 7s$:}
The $6d - 7s$ level crossing occurs in low ionisation stages of isoelectronic sequences with $N \geq 87$. For example in the Fr isoelectronic sequence, it is seen in Ac$^{2+}$~\cite{berengut11pra0}, which has $7p$, $6d$, and $5f$ orbitals all within optical range of the $7s$ ground state. This level crossing is available in isoelectronic sequences with $N \geq 87$. The $6f - 7s$ crossing should exist in all sequences with $N \geq 87$ since the $6f$ shell is never occupied. However the $6f$ orbital is at such high energy that the crossing occurs in very highly charged ions, with $Z > 100$. Therefore the crossing is not shown in \Fig{fig:ordering} since it will not occur in stable isotopes.

\section{$4f-5s$ crossing}
\label{sec:fs_crossing}

In this section, we examine the $4f - 5s$ crossing in greater detail. As mentioned previously, this level crossing occurs in ions with a relatively high degree of ionization. \Tref{tab:4f5sCI} presents CI calculations of some ions near this crossing with up to three valence electrons. As can be seen from the tables, the range of values for the charge on the ion \Zi\ for which the level crossing occurs remains relatively stable. This reinforces the general rule of thumb that given an ion near a level crossing, simultaneously increasing or decreasing both the charge $Z$ and the number of electrons $N$ by the same amount will result in another ion near the same level crossing.

The $4f - 5s$ level crossing is particularly unique in that it is the only available level crossing between levels of different parity in HCIs. This hints to the possibility of optical E1 transitions in these ions, which could be useful for cooling and trapping of HCIs. Two points are worth noting, however: firstly, in HCIs the strength of E1 transitions is suppressed compared  to near-neutral atoms (\Sec{sec:ejmjscaling}); and secondly with \mbox{$\Delta l = 3$} for an $s-f$ transition, it will tend to proceed via configuration mixing which greatly reduces its strength. Examples include $_{63}$Eu$^{14+}$ which has E1 transitions between the ground state, $4f^{2}5s$ ($J^P = 3.5^+$), and excited states $4f5s^{2}$ ($J = 2.5^-$) and $4f^3$ ($J = 4.5^-$), with energy intervals $17478~\text{cm}^{-1}$ and $28828~\text{cm}^{-1}$, respectively.

Perhaps the most interesting examples presented in \Tref{tab:4f5sCI} are the two-valence-electron ion Sm$^{14+}$, which was studied in~\cite{berengut10prl}, and the three-valence-electron ion Eu$^{14+}$. Both of these ions have ground states with half-open $5s$ shell, which means that both have optical $s-f$ and $f-s$ ground state transitions. The two E1 transitions in Eu$^{14+}$ mentioned previously are of this type, and it means that they will have $q$-values that are of opposite sign. On the other hand they may be too broad for high-precision clocks. Better reference transitions for clocks are strongly suppressed E1 transitions, suppressed M1, and E2 transitions.

It is also possible to have level crossings in hole states, where one or two electrons are removed from otherwise closed shells and effectively give rise to a similar structure as one- or two-valence-electron systems. The specific cases of Ir$^{16+}$ and Ir$^{17+}$ were studied in~\cite{berengut11prl} for the hole case, which leaves all intermediate cases. In general, intermediate ions with more than one electron result in large configuration spreading, significantly complicating the level structure of the ion. This is not true for hole cases, which allow for simpler level structures, yet providing the benefit of increased $Z$ and therefore high sensitivity to $\alpha$-variation.

\begin{longtable}{@{\extracolsep{19pt}}lclcc}
\caption{\label{tab:4f5sCI} Configuration interaction calculations for the level structure of highly charged ions with one, two or three valence electrons and $4f-5s$ intervals below $100\,000~\text{cm}^{-1}$. In general, the ellipses $\ldots$ used indicate that there are more fine-structure states available which we omit for brevity.}\\
\hline\hline
N & Ion & Config. & $J^P$ & Energy (cm$^{-1}$)\\
\hline
\endfirsthead
\caption{(continued)}\\
\hline\hline
N & Ion & Config. & $J^P$ & Energy (cm$^{-1}$)\\
\hline
\endhead
\hline \endfoot
\hline \hline \endlastfoot
47 & $_{60}$Nd$^{13+}$ &       $5s$     & 0.5$^+$ & 0 \\*
   &                   &       $4f$     & 2.5$^-$ & 64084 \\*
   &                   &       $4f$     & 3.5$^-$ & 68480 \\*
47 & $_{61}$Pm$^{14+}$ &       $5s$     & 0.5$^+$ & 0 \\*
   &                   &       $4f$     & 2.5$^-$ & 8902 \\*
   &                   &       $4f$     & 3.5$^-$ & 14290 \\*
47 & $_{62}$Sm$^{15+}$ &       $4f$     & 2.5$^-$ & 0 \\*
   &                   &       $4f$     & 3.5$^-$ & 6485 \\*
   &                   &       $5s$     & 0.5$^+$ & 51314 \\
48 & $_{60}$Nd$^{12+}$ &       $5s^{2}$ & 0$^+$ & 0\\*
   &                   & $4f5s$ & 2$^-$ & 86136\\*
   &                   & $4f5s$ & 3$^-$ & 87464\\*
   &                   & $4f5s$ & 4$^-$ & 90435\\*
   &                   & $4f5s$ & 3$^-$ & 96929\\
48 & $_{61}$Pm$^{13+}$ &       $5s^{2}$ & 0$^+$ & 0\\*
   &                   & $4f5s$ & 2$^-$ & 32742\\*
   &                   & $4f5s$ & 3$^-$ & 34261\\*
   &                   & $4f5s$ & 4$^-$ & 38030\\*
   &                   & $4f5s$ & 3$^-$ & 44299\\*
   &                   & $4f^{2}$       & 4$^+$ & 98912\\
48 & $_{62}$Sm$^{14+}$ & $4f5s$ & 2$^-$ & 0 \\*
   &                   & $4f5s$ & 3$^-$ & 1697\\*
   &                   & $4f5s$ & 4$^-$ & 6381\\*
   &                   & $4f^{2}$       & 4$^+$ & 11223\\*
   &                   & $4f5s$ & 3$^-$ & 12460\\*
   &                   & $4f^{2}$       & 5$^+$ & 16095\\*
   &                   & \ldots         &       & \\*
   &                   &       $5s^{2}$ & 0$^+$ & 25338\\*
   &                   & $4f^{2}$       & 4$^+$ & 31069\\*
   &                   & \ldots         &       & \\
48 & $_{63}$Eu$^{15+}$ & $4f^{2}$       & 4$^+$ & 0\\*
   &                   & $4f^{2}$       & 5$^+$ & 5886\\*
   &                   & \ldots         &       & \\*
   &                   & $4f5s$ & 2$^-$ & 48780\\*
   &                   & $4f5s$ & 3$^-$ & 50643\\*
   &                   & \ldots         &       & \\
49 & $_{62}$Sm$^{13+}$ & $4f5s^{2}$ & 2.5$^-$ & 0\\*
   &                   & $4f5s^{2}$ & 3.5$^-$ & 6189\\*
   &                   & $4f^{2}5s$ & 3.5$^+$ & 40211\\*
   &                   & $4f^{2}5s$ & 4.5$^+$ & 42454\\*
   &                   & \ldots         &         & \\
49 & $_{63}$Eu$^{14+}$ & $4f^{2}5s$ & 3.5$^+$ & 0\\*
   &                   & $4f^{2}5s$ & 4.5$^+$ & 2601\\*
   &                   & $4f^{2}5s$ & 5.5$^+$ & 6663\\*
   &                   & $4f^{2}5s$ & 1.5$^+$ & 10711\\*
   &                   & \ldots         &         & \\*
   &                   & $4f5s^{2}$ & 2.5$^-$ & 17478\\*
   &                   & $4f^{2}5s$ & 3.5$^+$ & 20722\\*
   &                   & \ldots         &       & \\*
   &                   & $4f5s^{2}$ & 3.5$^-$ & 24854\\*
   &                   & $4f^{3}$       & 4.5$^-$ & 28828\\*
   &                   & \ldots         &       & \\
49 & $_{64}$Gd$^{15+}$ & $4f^{3}$       & 4.5$^-$ & 0\\*
   &                   & $4f^{3}$       & 5.5$^-$ & 4768\\*
   &                   & $4f^{3}$       & 6.5$^-$ & 9711\\*
   &                   & $4f^{3}$       & 1.5$^-$ & 24137\\*
   &                   & \ldots         &         & \\*
   &                   & $4f^{2}5s$ & 3.5$^+$ & 30172\\*
   &                   & $4f^{3}$       & 4.5$^-$ & 31911\\*
   &                   & \ldots         &         & \\
49 & $_{65}$Tb$^{16+}$ & $4f^{3}$       & 4.5$^-$ & 0\\*
   &                   & $4f^{3}$       & 5.5$^-$ & 5702\\*
   &                   & $4f^{3}$       & 6.5$^-$ & 11527\\*
   &                   & $4f^{3}$       & 1.5$^-$ & 25637\\*
   &                   & \ldots         &         & \\*
   &                   & $4f^{2}5s$ & 3.5$^+$ & 94034\\*
   &                   & $4f^{2}5s$ & 4.5$^+$ & 97331\\
\end{longtable}

\section{$4f - 5p$ crossing}
\label{sec:fp_crossingOne}

The $4f - 5p$ crossing differs from the $4f - 5s$ crossing in that the orbitals are of the same parity and the $5p$ orbital has strong fine-structure splitting. HCIs near this crossing can have M1 transitions even without configuration mixing since the single-electron $5p_{3/2} - 4f_{5/2}$ transition is M1-allowed (although not in the non-relativistic limit since $\Delta l = 2$). Since the ratio of M1/E1 transition strengths is larger in HCIs relative to near-neutral ions (due to the suppression of E1 transitions), these ions can have rich physics to exploit in clocks. Additionally E2-allowed transitions are plentiful, and these can have linewidths which are more appropriate for reference transitions.
 
With the possibility of up to 14 electrons in the $f$-shell and 6 electrons in the $p$-shell, there are many ions that have this crossing, from single-valence-electron examples like $_{59}$Pr$^{10+}$ to the nineteen-valence-electron (single hole) $_{74}$W$^{7+}$. \Tref{tab:4f5pDF} presents Dirac-Fock calculations of energy levels in the $V^{N-1}$ approximation, with a $5p^x$ shell included above the closed Cd ($N=48$) core. The $5p^x$ shell ($x = N-1 - 48$) is included by weighting the potential of the filled $5p^6$ shell by the factor $x/6$. The position of the crossing from the Dirac-Fock estimate does not always agree with the configuration interaction calculation, but at least provides for a reasonable starting point.

\begin{table}[tb]
\label{tab:4f5pDF}
\caption{Weighted Dirac-Fock energy intervals calculated in the $V^{N-1}$ potential for highly charged ions near the $4f - 5p$ level crossing. The Dirac-Fock procedure includes a Cd core and a weighted $5p$ shell: [Kr] $5s^2 4d^{10} 5p^x$, with $x = N - 49$. }
\begin{ruledtabular}
\begin{tabular}{lccrrr}
$N$ & $x$ & Ion & \multicolumn{3}{c}{Energy relative to $4f_{5/2}$ orbital (cm$^{-1}$)}\\
&&& $5p_{1/2}$ & $5p_{3/2}$ & $4f_{7/2}$ \\
\hline
49& 0 & $_{57}$La$^{8+}$  & -114738 &  -86257 & 1698 \\
 && $_{58}$Ce$^{9+}$  &  -70791 &  -37066 & 2406 \\
 && $_{59}$Pr$^{10+}$ &  -20261 &   19230 & 3200 \\
 && $_{60}$Nd$^{11+}$ &   36280 &   82094 & 4085 \\
 && $_{61}$Pm$^{12+}$ &   98419 &  151157 & 5066 \\
50 & 1 & $_{60}$Nd$^{10+}$ & -130828 &  -82538 & 4013 \\
 && $_{61}$Pm$^{11+}$ &  -77060 &  -21675 & 4988 \\
 && $_{62}$Sm$^{12+}$ &  -17868 &   45257 & 6065 \\
 && $_{63}$Eu$^{13+}$ &   46466 &  118020 & 7249 \\
51 & 2 & $_{61}$Pm$^{10+}$ & -100727 &  -47479 & 4916 \\
 && $_{62}$Sm$^{11+}$ &  -43378 &   17463 & 5987 \\
 && $_{63}$Eu$^{12+}$ &   19191 &   88305 & 7164 \\
 && $_{64}$Gd$^{13+}$ &   86736 &  164841 & 8456 \\
52 & 3 & $_{61}$Pm$^{9+}$  & -123164 &  -72027 & 4849 \\
 && $_{62}$Sm$^{10+}$ &  -67686 &   -9103 & 5915 \\
 && $_{63}$Eu$^{11+}$ &   -6903 &   59797 & 7086 \\
 && $_{64}$Gd$^{12+}$ &   58917 &  134449 & 8370 \\
53 & 4 & $_{61}$Pm$^{8+}$  & -144342 &  -95288 & 4787 \\
 && $_{62}$Sm$^{9+}$  &  -90768 &  -34415 & 5849 \\
 && $_{63}$Eu$^{10+}$ &  -31796 &   32520 & 7014 \\
 && $_{64}$Gd$^{11+}$ &   32282 &  105269 & 8290 \\
54 & 5 & $_{62}$Sm$^{8+}$  & -112596 &  -58445 & 5788 \\
 && $_{63}$Eu$^{9+}$  &  -55462 &    6497 & 6948 \\
 && $_{64}$Gd$^{10+}$ &    6852 &   77323 & 8218 \\
 && $_{65}$Tb$^{11+}$ &   74101 &  153821 & 9606 \\
55 & 6 & $_{62}$Sm$^{7+}$  & -133139 &  -81160 & 5734 \\
 && $_{63}$Eu$^{8+}$  &  -77874 &  -18239 & 6889 \\
 && $_{64}$Gd$^{9+}$  &  -17346 &   50637 & 8153 \\
 && $_{65}$Tb$^{10+}$ &   48176 &  125243 & 9533 \\
\end{tabular}
\end{ruledtabular}
\end{table}

In \Tref{tab:4f5pCI} we present CI calculations for two- and three-valence-electron ions near the $4f - 5p$ crossing. Interesting examples here include the $_{60}$Nd$^{10+}$ ion which has a mixed $4f5p$ ground state from which narrow transitions are available to $4f^2$ and $5p^2$ configurations. These would have $q$ values of opposite sign, and so a clock using these transition would be a good probe of $\alpha$-variation. On the other hand configuration mixing ensures that the three-valence-electron ions $_{60}$Nd$^{9+}$ and $_{61}$Pm$^{10+}$ ion have good E2 and M1 transitions well within the range of usual optical and near-IR lasers. At the heavier end of the spectrum of ions which have this crossing are W$^{7+}$ and W$^{8+}$, with one and two holes in otherwise filled orbitals, respectively. These were studied in~\cite{berengut11prl}.

\begin{table}
\label{tab:4f5pCI}
\caption{Configuration interaction calculations for the level structure of HCIs with two or three valence electrons and $4f-5p$ intervals below $100\,000~\text{cm}^{-1}$. The ellipses $\ldots$ are used to indicate that there are more fine-structure states available which we omit for brevity.}
\begin{ruledtabular}
\begin{tabular}{lclcc}
N & Ion & Config. & J & Energy (cm$^{-1}$)\\
\hline
50 & $_{58}$Ce$^{8+}$  &       $5p^{2}$ & 0 & 0\\
   &                   &       $5p^{2}$ & 1 & 23362\\
   &                   &       $5p^{2}$ & 2 & 31033\\
   &                   & $4f5p$         & 3 & 92661\\
   &                   & $4f5p$         & 4 & 98806\\
50 & $_{59}$Pr$^{9+}$  &       $5p^{2}$ & 0 & 0\\
   &                   &       $5p^{2}$ & 1 & 28273\\
   &                   &       $5p^{2}$ & 2 & 34999\\
   &                   & $4f5p$         & 3 & 44738\\
   &                   & $4f5p$         & 4 & 51669\\
   &                   & $4f5p$         & 5 & 86593\\
50 & $_{60}$Nd$^{10+}$ & $4f5p$         & 3 & 0\\
   &                   & $4f5p$         & 2 & 3640\\
   &                   & $4f5p$         & 4 & 7701\\
   &                   &       $5p^{2}$ & 0 & 9060\\
   &                   & $4f^{2}$       & 5 & 33730\\
   &                   & $4f^{2}$       & 6 & 36668\\
   &                   &       $5p^{2}$ & 1 & 42578\\
51 & $_{59}$Pr$^{8+}$  &       $5p^{3}$       & 1.5 & 0\\
   &                   &       $5p^{3}$       & 1.5 & 26953\\
   &                   &       $5p^{3}$       & 2.5 & 34494\\
   &                   & $4f5p^{2}$       & 2.5 & 47413\\
   &                   & $4f5p^{2}$       & 3.5 & 50927\\
   &                   &       $5p^{3}$       & 0.5 & 51929\\
   &                   & $4f5p^{2}$       & 3.5 & 68470\\
   &                   & \ldots               &     & \\
51 & $_{60}$Nd$^{9+}$  & $4f5p^{2}$       & 2.5 & 0\\
   &                   & $4f5p^{2}$       & 3.5 & 6429\\
   &                   &       $5p^{3}$       & 1.5 & 10613\\
   &                   & $4f5p^{2}$       & 2.5 & 25124\\
   &                   & $4f5p^{2}$       & 3.5 & 27641\\
   &                   & \ldots               &     & \\
   &                   & $4f^{2}5p$       & 4.5 & 58361\\
   &                   & \ldots               &     & \\
51 & $_{61}$Pm$^{10+}$ & $4f5p^{2}$       & 2.5 & 0\\
   &                   & $4f^{2}5p$       & 4.5 & 3937\\
   &                   & $4f5p^{2}$       & 3.5 & 6992\\
   &                   & $4f^{2}5p$       & 3.5 & 9483\\
   &                   & $4f^{2}5p$       & 5.5 & 10844\\
   &                   & $4f^{2}5p$       & 2.5 & 13732\\
   &                   & $4f^{2}5p$       & 3.5 & 16000\\
   &                   & \ldots               &     & \\
   &                   & $4f5p^{2}$       & 2.5 & 31646\\
   &                   & \ldots               &     & \\
\end{tabular}
\end{ruledtabular}
\end{table}

\section{$5f - 6p$ crossing}
\label{sec:fp_crossingTwo}

The $5f - 6p$ level crossing is similar in many ways to the $4f - 5p$ crossing, with some important differences. Since this crossing occurs in ions with very high $Z$, the fine-structure splitting of the $6p$ levels is very large, and near the level crossing is usually much larger than the $5f - 6p$ interval. This provides advantages over the $4f - 5p$ crossing in that there are a larger number of ions available where one of these orbitals cross and also that the large fine-structure splitting causes a simplification of the level structure. In cases where the $6p_{1/2}$ and $5f$ levels cross, such as near Cf$^{17+}$, there is an enhancement of sensitivity to $\alpha$-variation~\cite{berengut12arxiv1}. The lower component of the $p_{1/2}$ Dirac spinor has an $s_{1/2}$ structure and is not small because of the high $Z$. This means that the $p_{1/2}$ orbital has a $q$-value comparable to an $s$-wave orbital.

\begin{table}[tb]
\caption{\label{tab:5f6pDF} Weighted Dirac-Fock energy intervals calculated in the $V^{N-1}$ potential for highly charged ions near the $5f - 6p$ level crossing. The Dirac-Fock procedure includes a Hg core ($N=80$) and a weighted $6p$ shell: [Xe] $6s^2 5d^{10} 4f^{14} 6p^x$, with $x = N - 81$.}
\begin{ruledtabular}
\begin{tabular}{lcrrr}
$N$ & Ion & \multicolumn{3}{c}{Energy relative to $5f_{5/2}$ orbital (cm$^{-1}$)} \\
&&$6p_{1/2}$ & $6p_{3/2}$ & $5f_{7/2}$\\
\hline
81 & $_{90}$Th$^{9+}$   & -182828 & -94604 & 4964 \\
81 & $_{91}$Pa$^{10+}$  & -167192 & -65878 & 6440 \\
81 & $_{92}$U$^{11+}$   & -148656 & -33232 & 8044 \\
81 & $_{93}$Np$^{12+}$  & -127546 &   3067 & 9775 \\
81 & $_{94}$Pu$^{13+}$  & -104132 &  42813 & 11636 \\
81 & $_{95}$Am$^{14+}$  &  -78639 &  85844 & 13630 \\
81 & $_{96}$Cm$^{15+}$  &  -51267 & 132033 & 15760 \\
81 & $_{97}$Bk$^{16+}$  &  -22195 & 181274 & 18030 \\
81 & $_{98}$Cf$^{17+}$  &    8415 & 233481 & 20446 \\
81 & $_{99}$Es$^{18+}$  &   40410 & 288583 & 23012 \\
81 & $_{100}$Fm$^{19+}$ &   73643 & 346516 & 25732 \\
82 & $_{95}$Am$^{13+}$  & -258013 & -86547 & 14560 \\
82 & $_{96}$Cm$^{14+}$  & -237735 & -47179 & 16727 \\
82 & $_{97}$Bk$^{15+}$  & -215633 &  -4627 & 19033 \\
82 & $_{98}$Cf$^{16+}$  & -191886 &  41006 & 21484 \\
82 & $_{99}$Es$^{17+}$  & -166663 &  89635 & 24085 \\
83 & $_{96}$Cm$^{13+}$  & -248665 & -63867 & 16320 \\
83 & $_{97}$Bk$^{14+}$  & -227341 & -22372 & 18614 \\
83 & $_{98}$Cf$^{15+}$  & -204330 &  22234 & 21052 \\
83 & $_{99}$Es$^{16+}$  & -179808 &  69859 & 23638 \\
84 & $_{96}$Cm$^{12+}$  & -259040 & -79947 & 15920 \\
84 & $_{97}$Bk$^{13+}$  & -238508 & -39525 & 18205 \\
84 & $_{98}$Cf$^{14+}$  & -216246 &   4043 & 20630 \\
84 & $_{99}$Es$^{15+}$  & -192434 &  50657 & 23202 \\
85 & $_{96}$Cm$^{11+}$  & -262798 & -89202 & 15076 \\
85 & $_{97}$Bk$^{12+}$  & -242718 & -49486 & 17332 \\
85 & $_{98}$Cf$^{13+}$  & -220875 &  -6600 & 19728 \\
85 & $_{99}$Es$^{14+}$  & -197458 &  39345 & 22269 \\
85 & $_{100}$Fm$^{15+}$ & -172644 &  88261 & 24960 \\
86 & $_{97}$Bk$^{11+}$  & -259178 & -71990 & 17413 \\
86 & $_{98}$Cf$^{12+}$  & -238450 & -30538 & 19816 \\
86 & $_{99}$Es$^{13+}$  & -216091 &  14017 & 22363 \\
86 & $_{100}$Fm$^{14+}$ & -192285 &  61579 & 25059 \\
87 & $_{97}$Bk$^{10+}$  & -268654 & -87268 & 17032 \\
87 & $_{98}$Cf$^{11+}$  & -248716 & -46899 & 19426 \\
87 & $_{99}$Es$^{12+}$  & -227101 &  -3390 & 21960 \\
87 & $_{100}$Fm$^{13+}$ & -204002 &  43149 & 24643 \\
87 & $_{101}$Md$^{14+}$ & -179597 &  92630 & 27480 \\
\end{tabular}
\end{ruledtabular}
\end{table}

\begin{table}
\caption{\label{tab:5f6pCI_2e} Configuration interaction estimates for the level structure of highly charged ions with two valence electrons and $5f-6p$ intervals below $100\,000~\text{cm}^{-1}$. Ellipses ($\ldots$) indicate that there are more fine-structure states that have been omitted. All levels have even parity.}
\begin{ruledtabular}
\begin{tabular}{lclcc}
N & Ion & Config. & J & Energy (cm$^{-1}$)\\
\hline
82 & $_{95}$Am$^{13+}$  &       $6p^{2}$ & 0 & 0\\
   &                    & $5f6p$ & 3 & 89786\\
   &                    & $5f6p$ & 2 & 97898\\
82 & $_{96}$Cm$^{14+}$  &       $6p^{2}$ & 0 & 0\\
   &                    & $5f6p$ & 3 & 63664\\
   &                    & $5f6p$ & 2 & 72221\\
   &                    & $5f6p$ & 3 & 83564\\
   &                    & $5f6p$ & 4 & 85846\\
82 & $_{97}$Bk$^{15+}$  &       $6p^{2}$ & 0 & 0\\
   &                    & $5f6p$ & 3 & 36004\\
   &                    & $5f6p$ & 2 & 44444\\
   &                    & $5f6p$ & 3 & 58033\\
   &                    & $5f6p$ & 4 & 59702\\
   &                    &       $5f^{2}$ & 4 & 90788\\
82 & $_{98}$Cf$^{16+}$  &       $6p^{2}$ & 0 & 0\\
   &                    & $5f6p$ & 3 & 7452\\
   &                    & $5f6p$ & 2 & 14775\\
   &                    &       $5f^{2}$ & 4 & 28824\\
   &                    & $5f6p$ & 3 & 31436\\
   &                    &       $5f^{2}$ & 4 & 36157\\
   &                    & \ldots         &   & \\
82 & $_{99}$Es$^{17+}$  &       $5f^{2}$ & 4 & 0\\
   &                    &       $5f^{2}$ & 2 & 4536\\
   &                    & $5f6p$ & 3 & 5323\\
   &                    &       $5f^{2}$ & 5 & 20460\\
   &                    &       $5f^{2}$ & 4 & 21371\\
   &                    & $5f6p$ & 2 & 22436\\
   &                    &       $6p^{2}$ & 0 & 22871\\
   &                    &       $5f^{2}$ & 3 & 24795\\
   &                    & $5f6p$ & 3 & 35959\\
   &                    &       $5f^{2}$ & 6 & 42232\\
   &                    & \ldots         &   & \\
82 & $_{100}$Fm$^{18+}$ &       $5f^{2}$ & 4 & 0\\
   &                    &       $5f^{2}$ & 2 & 9828\\
   &                    &       $5f^{2}$ & 5 & 22772\\
   &                    &       $5f^{2}$ & 4 & 27627\\
   &                    &       $5f^{2}$ & 3 & 28876\\
   &                    & $5f6p$ & 3 & 37642\\
   &                    &       $5f^{2}$ & 6 & 39897\\
   &                    & \ldots         &   & \\
   &                    & $5f6p$ & 3 & 67819\\
   &                    & $5f6p$ & 4 & 73676\\
   &                    & \ldots         &   & \\
\end{tabular}
\end{ruledtabular}
\end{table}

\begin{table}[!tb]
\caption{\label{tab:5f6pCI_3e} Configuration interaction calculations for the level structure of HCIs with three valence electrons and $5f-6p$ intervals below $100\,000~\text{cm}^{-1}$. Ellipses ($\ldots$) indicate that there are more fine-structure states that have been omitted. All levels have odd parity.}
\begin{ruledtabular}
\begin{tabular}{lclcc}
N & Ion & Config. & J & Energy (cm$^{-1}$)\\
\hline
83 & $_{92}$U$^{9+}$    & $6p^3$     & 1.5 & 0\\
   &                    & $5f6p^{2}$ & 2.5 & 70210\\
   &                    & $5f6p^{2}$ & 3.5 & 82945\\
83 & $_{96}$Cm$^{13+}$  & $5f6p^{2}$ & 2.5 & 0\\
   &                    & $5f6p^{2}$ & 3.5 & 18815\\
   &                    & $5f^{2}6p$ & 4.5 & 83815\\
   &                    & $5f^{2}6p$ & 2.5 & 97251\\
   &                    & $5f^{2}6p$ & 3.5 & 97765\\
83 & $_{97}$Bk$^{14+}$  & $5f6p^{2}$ & 2.5 & 0\\
   &                    & $5f6p^{2}$ & 3.5 & 20858\\
   &                    & $5f^{2}6p$ & 4.5 & 58127\\
   &                    & $5f^{2}6p$ & 2.5 & 72447\\
   &                    & $5f^{2}6p$ & 3.5 & 73189\\
   &                    & $5f^{2}6p$ & 1.5 & 76551\\
   &                    & $5f^{2}6p$ & 5.5 & 79687\\
   &                    & \ldots         &   & \\
83 & $_{98}$Cf$^{15+}$  & $5f6p^{2}$ & 2.5 & 0\\
   &                    & $5f6p^{2}$ & 3.5 & 22742\\
   &                    & $5f^{2}6p$ & 4.5 & 31188\\
   &                    & $5f^{2}6p$ & 2.5 & 46699\\
   &                    & $5f^{2}6p$ & 3.5 & 47136\\
   &                    & $5f^{2}6p$ & 1.5 & 49751\\
   &                    & $5f^{2}6p$ & 5.5 & 54895\\
   &                    & \ldots         &   & \\
83 & $_{99}$Es$^{16+}$  & $5f6p^{2}$ & 2.5 & 0\\
   &                    & $5f^{2}6p$ & 4.5 & 4928\\
   &                    & $5f^{2}6p$ & 3.5 & 19106\\
   &                    & $5f^{2}6p$ & 1.5 & 22246\\
   &                    & $5f^{2}6p$ & 2.5 & 23262\\
   &                    & $5f6p^{2}$ & 3.5 & 26967\\
   &                    & $5f^{2}6p$ & 5.5 & 30767\\
   &                    & \ldots         &   & \\
   &                    &       $5f^{3}$ & 5.5 & 55606\\
   &                    & \ldots         &   & \\
   &                    &       $5f^{3}$ & 6.5 & 64091\\
   &                    &       $5f^{3}$ & 1.5 & 65019\\
   &                    & \ldots         &   & \\
83 & $_{100}$Fm$^{17+}$ & $5f^{2}6p$ & 4.5 & 0\\
   &                    &       $5f^{3}$ & 4.5 & 8162\\
   &                    & $5f^{2}6p$ & 2.5 & 11213\\
   &                    & $5f^{2}6p$ & 1.5 & 12028\\
   &                    & $5f^{2}6p$ & 3.5 & 18134\\
   &                    &       $5f^{3}$ & 5.5 & 22763\\
   &                    & $5f^{2}6p$ & 2.5 & 28805\\
   &                    & \ldots         &   & \\
   &                    &       $5f^{3}$ & 1.5 & 33451\\
   &                    & \ldots         &   & \\
\end{tabular}
\end{ruledtabular}
\end{table}

In \Tref{tab:5f6pDF} we present weighted Dirac-Fock orbital energies for ions near the $5f-6p$ crossing in the $V^{N-1}$ approximation. For the single-valence electron case ($N=81$) two crossings are seen. The first occurs between U$^{11+}$ and Np$^{12+}$ and corresponds to the $5f - 6p_{3/2}$ crossing, while the $5f - 6p_{1/2}$ crossing occurs near Cf$^{17+}$. This second crossing is only shown in \Tref{tab:5f6pDF} for $N=81$ because it is soon pushed to ions with $Z>100$ (although clearly it will still occur for two- or three-valence-electron ions).

Configuration interaction calculations for some interesting HCIs with the $5f-6p$ crossing are shown in \Tref{tab:5f6pCI_2e} (two-valence-electron ions) and \Tref{tab:5f6pCI_3e} (three-valence-electron ions). As with the ions near the $4f-5p$ level crossing, many M1 and E2 transitions are available within the optical range corresponding to single-electron $p-f$ transitions. Clearly the difficulty with exploiting this crossing is that many of the elements with transitions near it are not stable and do not occur naturally. In \cite{berengut12arxiv1} we studied Cf$^{16+}$ in some detail since it is relatively stable (with isotopes that live up to several hundred years) and has the $6p_{1/2}-5f$ crossing mentioned previously. Using hole transitions is not possible with this crossing since there would need to be around 14 or 15 valence electrons (corresponding to the crossing of filled $6p_{1/2}^2$ and $5f^{14}$ shells, minus one or two electrons), and this would require $Z>100$, well past the somewhat stable elements.

An interesting example that makes use of the $5f - 6p_{3/2}$ crossing is the three-valence-electron U$^{9+}$. Because of the large fine-structure splitting, the first two valence electrons fill the $6p_{1/2}^2$ subshell. The third valence electron is in the $6p_{3/2}$ subshell (ground state) but may be excited to the $5f_{5/2}$ and $5f_{7/2}$ orbitals. These transitions are shown in \Tref{tab:5f6pCI_3e}. The transitions (M1 at $70210\ \cm$ and E2 at $82945\ \cm$) will not be particularly sensitive to $\alpha$-variation, but $^{235}$U is interesting also because it has a 76 eV nuclear transition which may soon come within XUV laser range~\cite{cingoz12nat}. The nuclear transition would have high sensitivity to variation of fundamental constants, and the electronic transition could then form an `anchor' (relatively insensitive) transition.

\section{Approximate Scaling Laws}
\label{sec:scalings}

It is useful to have simple estimates of various properties of HCIs given existing knowledge of an appropriate neutral or near neutral ion. In the following sections we derive approximate scaling laws for highly charged ions that state how a particular property will change along an isoelectronic sequence with increasing $Z$. Our approach is similar to that of~\cite{gillaspy01jpb}, however we use the formalism of effective charges which we believe will be more useful to experimentalists. A summary of our results is presented in \Tref{tab:scalings}.

\subsection{Coefficients of Linear Fitting for Effective Charge}

Recall the approximate formula for the non-relativistic energy of an electron in a screened Coulomb potential, $V(r) \sim -{Z_a}/{r}$,
\begin{equation}
 E_n = - \frac{(\Zi + 1)^2}{2\nu^2} = -\frac{\Za^2}{2n^2}\label{eqn:nonrelenergy} \,,
\end{equation}
where $n$ is the integer principal quantum number and $\nu$ is the effective principal quantum number which is introduced to keep agreement with experimentally observed energies. For the purposes of this work on highly charged ions, it is more convenient introduce an effective charge \Za\ as an alternative to $\nu$ that represents the (non-integer) effective screened charge of the potential that the external electron `sees'. In this formulation $n$ is kept as the usual integer principal quantum number. \Za\ scales nearly linearly along an isoelectronic sequence as $Z$ (or equivalently \Zi) increases.

\begin{figure}[tb]
\caption{Calculated effective charge $\Za = \sqrt{\left|2 n^2 E\right|}$ (circles) versus ion charge \Zi\ for a valence $4f$ electron above a closed shell [Xe]\,$6s^2$ ($N=56$) core. The lines represent linear fits using the values tabulated in \Tref{tab:Zeffcoeffs} for the appropriate regions of $\Zi$.}
\label{fig:4fza}
\includegraphics[width=0.45\textwidth]{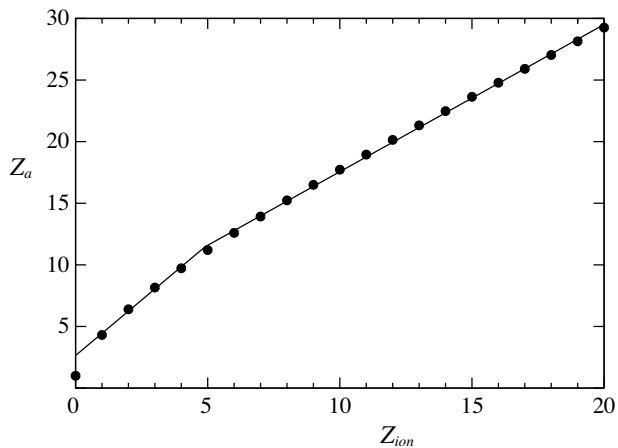}
\end{figure}

\Za\ can easily be calculated using Dirac-Fock energies for any ion. We also present fitting laws for \Za\ as a function of \Zi\ for several valence orbitals in \Tref{tab:Zeffcoeffs} that may be used to quickly obtain \Za. The data were obtained from one-valence-electron Dirac-Fock calculations, and we fit for the linear coefficients $A$ and $B$ according to
\begin{equation}
 \Za = A\,\Zi + B \,.
\end{equation}
Values for $A$ and $B$ are presented in \Tref{tab:Zeffcoeffs}. For large \Zi\ ($5 \leq \Zi \leq 20$, labelled H in \Tref{tab:Zeffcoeffs}), our calculations show that the linear approximation used above is in very good agreement with the calculated trend (see \Fig{fig:4fza}). Often experimental data is available for neutral or near-neutral ions and in order to extrapolate from these ions to HCIs requires a reasonable estimate of \Za\ for these ions. Therefore we also present fits across the domain \mbox{$1 \leq \Zi \leq 4$} (labelled L in \Tref{tab:Zeffcoeffs}) and values for the neutral atoms \mbox{$\Zi = 0$} (labelled N in \Tref{tab:Zeffcoeffs}). 

\begin{table}[htb]
\caption{Table of coefficients $A$ and $B$ for the effective charge $\Za = A\Zi + B$. For the regime column, L means the coefficients are more suitable for ions with $1 \leq \Zi \leq 4$ and H means that the coefficients are tailored for high ion charge $5 \leq \Zi \leq 20$, while N is the special case of neutral atoms where $\Zi = 0$. These values were tabulated using the Dirac-Fock energies of singly-occupied electron orbitals above closed shells.}
\label{tab:Zeffcoeffs}
\begin{ruledtabular}
\begin{tabular}{rcccc}
Orbital & N & Regime & $A$ & $B$\\
\hline
$4s$ & 19 & L & 1.233280 & 2.4530\\
     &    & H & 1.060127 & 3.3022\\
     &    & N &          & 2.1725\\
$5s$ & 37 & L & 1.402867 & 3.0221\\
     &    & H & 1.130601 & 4.4015\\
     &    & N &          & 2.6390\\
$6s$ & 55 & L & 1.542243 & 3.4892\\
     &    & H & 1.198431 & 5.2507\\
     &    & N &          & 3.0283\\
$7s$ & 87 & L & 1.750825 & 4.1439\\
     &    & H & 1.325290 & 6.3323\\
     &    & N &          & 3.5841\\
$4p$ & 31 & L & 1.362347 & 2.8994\\
     &    & H & 1.107311 & 4.1724\\
     &    & N &          & 2.5049\\
$5p$ & 49 & L & 1.525760 & 3.5762\\
     &    & H & 1.179904 & 5.3294\\
     &    & N &          & 3.0751\\
$6p$ & 81 & L & 1.774450 & 4.4369\\
     &    & H & 1.318796 & 6.7611\\
     &    & N &          & 3.7917\\
$3d$ & 21 & L & 1.188580 & 4.0655\\
     &    & H & 1.049841 & 4.7758\\
     &    & N &          & 3.9083\\
$4d$ & 39 & L & 1.447306 & 2.8779\\
     &    & H & 1.119397 & 4.5226\\
     &    & N &          & 2.3191\\
$5d$ & 71 & L & 1.724165 & 3.4661\\
     &    & H & 1.227029 & 5.9896\\
     &    & N &          & 2.6498\\
$4f$ & 57 & L & 1.801169 & 2.6428\\
     &    & H & 1.194715 & 5.6147\\
     &    & N &          & 1.0082\\
$5f$ & 89 & L & 2.028852 & 2.4084\\
     &    & H & 1.276494 & 6.1027\\
     &    & N &          & 1.2579
\end{tabular}
\end{ruledtabular}
\end{table}

Our approach is similar to that of Slater~\cite{slater30pr} for calculating the effective charge of electrons with shielding and the work that followed it. We see that the effective charge \Za\ is always bigger than $Z_i + 1$ (this correspondingly leads to $\nu < n$ for the other convention). This is because electrons spend a non-zero amount of time closer to the nucleus, and during that time experience a larger ion charge. It is worth noting these two schemes are equivalent for the $4f$ electron in neutral La, which experiences an effective charge very close to $Z_i + 1 = 1$ because it is much further from the nucleus than the $s$, $p$ and $d$ electrons below it. Therefore the use of $Z_i + 1$ gives similar results to \Za\ for electrons that are far removed from the potential of other electrons.

\subsection{Scaling of EJ and MJ Matrix Elements}
\label{sec:ejmjscaling}

In this section we present analytical estimates for the scaling of the EJ and MJ transition matrix elements. We use the following formulae to calculate both the non-relativistic electric and magnetic multipole reduced matrix elements (a relativistic treatment gives the same results). In the equations below, we seek to retain only the dependence of \Za\ in the relevant formulae whenever possible. The  E1 matrix element is 
\begin{equation}
 \langle nl | r | n'l' \rangle = \int P_{nl} r P_{n'l'} dr
\end{equation}
where the radial wavefunction $P_{nl}$ far away from the core electrons is
\begin{eqnarray}
 P_{nl} & = & N_{nl} \left( \frac{2Z_ar}{n} \right)^{l+1} e^{-\frac{Z_ar}{n}}F\left( -n+l+1, 2l+2, \frac{2Z_ar}{n} \right)\nonumber\\
 N_{nl} & = & \frac{1}{n(2l+1)!} \sqrt{\frac{Z_a(n+l)!}{(n-l-1)!}}\nonumber\ .
\end{eqnarray}
This allows the \Za\ dependence of the E1 integral to be calculated as
\begin{equation}
\left(\int_0^\infty r P_{i}P_{j} dr\right) \sim \left(Z_a\right)^{-1}\label{eqn:nonrelintegral}\ .
\end{equation}
The non-relativistic M1 matrix element does not scale with charge as it is a function of the angular momenta. Therefore, while the E1 matrix element decreases for a decrease in \Za, the M1 matrix element remains constant. In comparing highly-charged ions where (large \Za) and near neutral ions (small \Za), we see that M1 transitions can be as strong as E1 transitions as the latter decreases with increasing \Za. A similar treatment was adopted in~\cite{bates49ptrsa}, with effective principal quantum number $\nu$ (labelled $n^*$ in their equations) instead of effective charge \Za. For higher multipoles one obtains higher powers of the Coulomb radius
\begin{equation}
\label{eq:coulomb_radius}
\left< r^n \right> \sim \left( \frac{a_B}{\Za} \right)^n
\end{equation}
where $a_B$ is the Bohr radius, so that the general scaling law for EJ and MJ matrix elements is
\begin{equation}
\langle \kappa_i || q_J^{(E)} || \kappa_j \rangle \sim  \left(Z_a\right)^{-J}
\end{equation}
and
\begin{equation}
\langle \kappa_i || q_J^{(M)} || \kappa_j \rangle  \sim  \left(Z_a\right)^{1-J}\ .
\end{equation}
In general, E(J+1) matrix elements have the same \Za\ scaling as MJ matrix elements.

\subsection{Scaling of Polarizability and Blackbody Radiation Shift}

The blackbody radiation shift (BBR) for an adiabatic system can be calculated using the formula
\begin{equation}
\delta E = -\frac{1}{2}(831.9~\text{V/m})^2 \left( \frac{T(K)}{300} \right)^4 \alpha_0 (1+ \eta)
\end{equation}
where $\alpha_0$ is the static dipole polarizability and and $\eta$ is a small dynamic correction due to the frequency distribution, which for the purposes of this estimate we will disregard. The valence scalar polarizability of an atom in a state $v$ can be expressed as a sum over all excited intermediate states $n$ allowed by E1 selection rules.
\begin{equation}
\alpha_0 = \frac{2}{3(2j_v + 1)} \sum_n \frac{\langle v || r || n \rangle \langle n || r || v \rangle}{E_n - E_v}
\end{equation}
We showed in \Eref{eqn:nonrelintegral} that the reduced matrix element $\langle v || r || n \rangle$ scales simply as $1/\Za$. Also, the dependence of the non-relativistic energy on $Z_a$ is given by \Eref{eqn:nonrelenergy} to be $Z_a^2$. Therefore all terms in the summation have the same dependence on $Z_a$ and the total dependence on $Z_a$ must necessarily be the same. We must then have
\begin{equation}
\delta E \sim \alpha_0 \sim \left(\frac{1}{Z_a}\right)^4 \label{eqn:bbrreduction}\ .
\end{equation}
\Eref{eqn:bbrreduction} suggests that in systems with high effective charge (large \Za) such as highly charged ions, the BBR shift will be strongly suppressed as compared to neutral systems.

\subsection{Scaling of the Hyperfine structure}

Operators with large negative powers of radius will not follow the Coulomb radius scaling, \Eref{eq:coulomb_radius}, since the wavefunction at small distances cannot be described by $P_{nl}$. Instead we must use the approach of Fermi-Segr\'e~(see, e.g.~\cite{foldy58pr}) where the normalised squared wavefunction at the origin $\sim Z(\Zi+1)^2/\nu^3$. Since $\nu = n(\Zi+1)/\Za$, we then come to the following scaling law for the hyperfine $A$ coefficient:
\[
\frac{A}{g_I} \sim \frac{Z\Za^3}{(\Zi+1)}
\]
where we have factored out the nuclear $g$-factor $g_I$ which varies greatly between nuclei. We compare this scaling law with experimental data in \Tref{tab:scalingcomp}. A similar result may be derived for the electric quadrupole hyperfine constant $B$. We should also point out that the widths of hyperfine transitions will scale as $\omega^3 \sim A^3$, therefore relaxation of hyperfine structure will occur much faster in HCIs.

\begin{table}[tb]
\caption{Magnetic dipole hyperfine coefficients $A$ (calculated in~\cite{wu07chinesephys}) and their scaling with increasing $Z$ along the lithium isoelectronic sequence. Values of $Z_a$ were obtained from Dirac-Fock calculations using the relation $Z_a = \sqrt{|2n^2E|}$ where $n$ is the principal quantum number and $E$ is the orbital energy in atomic units. The notation $|_\text{p}$ means to use the values in the previous row of the table.}
\label{tab:scalingcomp}
\begin{ruledtabular}
\begin{tabular}{lcccccc}
Isotope & $I$ & $g_I$ & $A$~(MHz)~\cite{wu07chinesephys} & $Z_a$ & $\frac{(A/g_I)}{(A/g_I)|_\text{p}}$ & $\frac{\frac{ZZ^3_a}{Z_i + 1}}{\frac{ZZ^3_a}{Z_i + 1}\big|_{\text{p}}}$\\
\hline
$_3^7$Li          & 3/2 &  2.1709 &   399.34 & 1.25 &      &      \\
$_4^9$Be$^{+}$    & 3/2 & -0.7850 &  -625.55 & 2.31 & 4.35 & 4.20 \\
$_{\ 5}^{11}$B$^{2+}$ & 3/2 &  1.7924 &  3603.77 & 3.33 & 2.53 & 2.49 \\
$_{\ 6}^{13}$C$^{3+}$ & 1/2 &  1.4048 &  5642.40 & 4.35 & 2.00 & 2.00 \\ 
$_{\ 7}^{15}$N$^{4+}$ & 1/2 & -0.5664 & -3973.68 & 5.36 & 1.74 & 1.74 \\
$_{\ 8}^{17}$O$^{5+}$ & 5/2 & -0.7575 & -8474.13 & 6.36 & 1.59 & 1.59 \\
$_{\ 9}^{19}$F$^{6+}$ & 1/2 &  5.2578 & 88106.93 & 7.37 & 1.50 & 1.50
\end{tabular}
\end{ruledtabular}
\end{table}

\begin{table}[tb]
\caption{\label{tab:scalings} Scaling dependences for HCIs for various sources of systematic shifts in optical clocks.}
\begin{ruledtabular}
\begin{tabular}{ll}
$2^\textrm{nd}$ order Stark shift & $\sim 1/\Za^4$ \\
Blackbody shift & $\sim 1/\Za^4$ \\
$2^\textrm{nd}$ order Zeeman shift & suppressed\footnotemark[1] \\
Electric quadrupole shift & $\sim 1/Z_a^2$ \\
Fine-structure & $\sim Z^2\Za^3/(\Zi+1)$ \\
Hyperfine $A$ coefficient & $\sim Z\Za^3/(\Zi+1)$ \\
\end{tabular}
\end{ruledtabular}
\footnotetext[1]{The Zeeman shift is sensitive to the specific fine- and hyperfine-structure of the transition, but may be suppressed in HCIs due to a larger energy denominator.}
\end{table}

\section{Conclusion}

In this \paper, we have discussed all level crossings available in the periodic table and their characteristics. We separately discussed and identified several highly charged ions near level crossings and presented estimates for the energy intervals in some of these ions.  We also calculated scaling laws in terms of the effective screened charge \Za\ for transition matrix elements, energy intervals (including fine-structure), blackbody radiation and the hyperfine shift -- these provide a quick and reliable way to estimate size of these atomic properties given knowledge of these properties in a near-neutral ion. In order to facilitate these estimates, we have also tabulated empirical values for \Za\ for singly occupied electron orbitals above closed shells. Our scaling laws predict the BBR shifts in HCIs will be strongly suppressed. On the other hand, the hyperfine structure is much more important.

The potential future applications of HCIs as discussed in this \paper\ are clear -- the strong dependence of transitions in HCIs on the variation of the fine-structure constant makes them good candidates for laboratory tests of cosmological $\alpha$-variation. Finally, the existence of level crossings leads to the availability of transitions that can be excited by optical lasers. This is an experimental advantage HCIs have over nuclear clocks which have also been proposed to probe the variation of fundamental constants~\cite{peik03epl,campbell12prl}, but require lasers operating in the petahertz range to excite neutrons or protons.

\begin{acknowledgments}
This work was supported in part by the Australian Research Council. Supercomputer time was provided by an award under the Merit Allocation Scheme on the NCI National Facility at the Australian National University.
\end{acknowledgments}

\appendix*

\section{Deviations from Madelung Filling}
\label{app:madelung_deviations}

In some neutral atoms there are deviations from the Madelung order of filling; for example, these deviations are commonly observed in the transition elements. Similar to our treatment of the level crossings due to the Coulomb degeneracy, we now examine all available deviations from the Madelung filling order and characterise them. Recall that for level crossings that occur due to Coulomb degeneracy, we had crossings of type $s-f$, $s-d$ and $p-f$ only. In contrast, we find only find $s-d$ and $d-f$ type crossings in isoelectronic sequences starting from atoms that have deviations (see \Tref{tab:Deviations} for an exhaustive list). 

A specific example is lanthanum, with a ground state of [Xe] $5d 6s^{2}$. Here the $5d$ orbital is filled before the $4f$ orbital, while from the Madelung rule we would expect the $4f$ to be filled first. Because $5d$ has higher $n$ than $4f$, further along the isoelectronic sequence Coulomb degeneracy will cause the $4f$ orbital to be lower in energy than the $5d$ orbital. In Ce$^+$ the ground state is [Xe]$4f5d^2$. Pr$^{2+}$ has a ground state of [Xe]$4f^{3}$ which shows that all crossings have occurred by $\Zi=2$. We find that all crossings caused by the deviation from the Madelung rule occur at low ion charge.

\begin{table}[tb]
\caption{Deviations from Madelung filling (usual periodic table filling) in neutral atoms.}
\label{tab:Deviations}
\begin{ruledtabular}
\begin{tabular}{lll}
\multicolumn{1}{c}{Element} & \multicolumn{1}{c}{Actual Filling} & \multicolumn{1}{c}{Madelung Filling} \\
\hline
$_{24}$Cr & [Ar] $3d^{5} 4s$ & [Ar] $3d^{4} 4s^{2}$ \\
$_{29}$Cu & [Ar] $3d^{10} 4s$ & [Ar] $3d^{9} 4s^{2}$ \\
$_{41}$Nb & [Kr] $4d^{4} 5s$ & [Kr] $4d^{3} 5s^{2}$ \\
$_{42}$Mo & [Kr] $4d^{5} 5s$ & [Kr] $4d^{4} 5s^{2}$ \\
$_{44}$Ru & [Kr] $4d^{7} 5s$ & [Kr] $4d^{6} 5s^{2}$ \\
$_{45}$Rh & [Kr] $4d^{8} 5s$ & [Kr] $4d^{7} 5s^{2}$ \\
$_{46}$Pd & [Kr] $4d^{9} 5s$ & [Kr] $4d^{8} 5s^{2}$ \\
$_{47}$Ag & [Kr] $4d^{10} 5s$ & [Kr] $4d^{9} 5s^{2}$ \\
$_{57}$La & [Xe] $5d 6s^{2}$ & [Xe] $4f 6s^{2}$ \\
$_{58}$Ce & [Xe] $4f 5d 6s^{2}$ & [Xe] $4f^{2} 6s^{2}$ \\
$_{64}$Gd & [Xe] $4f^{7} 5d 6s^{2}$ & [Xe] $4f^{8} 6s^{2}$ \\
$_{78}$Pt & [Xe] $4f^{14} 5d^{9} 6s$ & [Xe] $4f^{14} 5d^{8} 6s^{2}$ \\
$_{79}$Au & [Xe] $4f^{14} 5d^{10} 6s$ & [Xe] $4f^{14} 5d^{9} 6s^{2}$ \\
$_{89}$Ac & [Rn] $6d 7s^{2}$ & [Rn] $5f 7s^{2}$ \\
$_{90}$Th & [Rn] $6d^{2} 7s^{2}$ & [Rn] $5f^{2} 7s^{2}$ \\
$_{91}$Pa & [Rn] $5f^{2} 6d 7s^{2}$ & [Rn] $5f^{3} 7s^{2}$ \\
$_{92}$U & [Rn] $5f^{3} 6d 7s^{2}$ & [Rn] $5f^{4} 7s^{2}$ \\
$_{93}$Np & [Rn] $5f^{4} 6d 7s^{2}$ & [Rn] $5f^{5} 7s^{2}$ \\
$_{96}$Cm & [Rn] $5f^{7} 6d 7s^{2}$ & [Rn] $5f^{8} 7s^{2}$ \\
\end{tabular}
\end{ruledtabular}
\end{table}

\setcounter{paragraph}{0}
\paragraph{$3d-4s$:}
This crossing occurs due to the additional stability offered by half-filled and filled $3d$ orbitals in Cr and Cu respectively. The half or complete filling of the $3d$ orbitals is preferred to a filled $4s$ orbital, as a result one of the $4s$ electrons in these atoms fill a $3d$ orbital instead. For Cr isoelectronic sequence, the remaining $4s$ electron eventually fills a $3d$ state instead.

\paragraph{$4d-5s$:}
In the ground states of Nb, Mo, Ru, Rh, Pd and Ag, the $4d$ shell fills before the $5s$ shell is closed. This is consistent with the calculations done on the Rb isoelectronic sequence, which reveals that the $4d$ and $5s$ orbitals cross at $Z = 39$.

\paragraph{$4f-5d$:}
Due to angular momentum and parity considerations, there exists optical E1 transitions in neutral La, Ce and Gd, as well as for near neutral ions in the vicinity of these atoms. The $4f$ and $5d$ orbitals must necessarily cross due to Coulomb degeneracy. Our calculations
show that this crossing occurs at $Z = 58$ for the La isoelectronic sequence.

\paragraph{$5d-6s$:}
The ground states of Pt and Au show a deviation from Madelung filling. According to our calculations, the $5d$ orbital lies below the $6s$ orbital for $Z = 56$, the next ion in the isoelectronic sequence. The crossing occurs near the beginning of the isoelectronic sequence because the orbitals are very close in energy to begin with.

\paragraph{$5f-6d$:}
The ground states of Ac, Th, Pa, U, Np and Cm have a single electron in the $6d_{3/2}$ orbital. In the example of neutral Th the $6d$ orbitals lie below the $5f$ orbitals, but in singly ionized Pa$^{+}$, the level crossing has occured~\cite{blaise92itsc} and the $5f$ orbitals lie around $\sim 5 000~\text{cm}^{-1}$ below the $6d$ orbitals. This crossing is also present in the Th$^{3+}$ ion that has several potential atomic clock transitions with enhanced sensitivity to $\alpha$-variation~\cite{flambaum09pra}.

\bibliography{references}

\end{document}